\documentclass[a4paper,reqno,11pt]{amsart}

\usepackage[utf8]{inputenc}
\usepackage[T1]{fontenc}
\usepackage{lmodern}
\usepackage[UKenglish]{babel}
\usepackage[UKenglish]{isodate}
\usepackage{amsmath, amsfonts, amssymb, amsthm, mathtools, cancel}
\usepackage{geometry}
\usepackage{braket}
\usepackage{mathrsfs}
\usepackage{xcolor}
\usepackage{hyperref}
\hypersetup{colorlinks = true}
\hypersetup{linktocpage} %to only link the page numbers and not the entire table of contents
\usepackage[capitalise]{cleveref} % to allow the format of cross-references to be determined automatically according to the “type” of cross-reference (equation, section, etc.) (capitalise = capitalization of all label names throughout the document)
\numberwithin{equation}{section}

\theoremstyle{plain}
\newtheorem{thm}{Theorem}[section]
\newtheorem{prop}[thm]{Proposition}

\newtheorem{lem}[thm]{Lemma}

\theoremstyle{definition}

\theoremstyle{remark}
\newtheorem{rmkx}[thm]{Remark}

  \newenvironment{rmk}
  {%
   \pushQED{\qed}\begin{rmkx}}
  {\popQED\end{rmkx}}

\newcommand{\interval}[4]{\mathopen{#1}#2
	\mathclose{}\mathpunct{},#3
	\mathclose{#4}}
\newcommand{\intcc}[2]{\interval{[}{#1}{#2}{]}}

\newcommand{\intoo}[2]{\interval{(}{#1}{#2}{)}}

\newcommand{\smallo}[1]{o\mathopen{}\left(#1\right)}

\newcommand{\cL}{\mathcal{L}}
\newcommand{\cM}{\mathcal{M}}
\newcommand{\cE}{\mathcal{E}}
\newcommand{\cN}{\mathcal{N}}

\newcommand{\cF}{\mathscr{F}^{\leq N}_+}

\newcommand{\hc}{\mbox{h.c.}}

\newcommand{\abs}[1]{\left\lvert#1\right\rvert}
\newcommand{\norm}[1]{\left\lVert#1\right\rVert}
\newcommand{\scalprod}[2]{\left\langle#1,#2\right\rangle}

\newcommand{\setst}[2]{\left\{#1\mathrel{}\middle|\mathrel{}#2\right\}}

\newcommand{\numberset}[1]{\mathbb{#1}}
\newcommand{\N}{\numberset{N}}
\newcommand{\Z}{\numberset{Z}}

\newcommand{\R}{\numberset{R}}
\newcommand{\C}{\numberset{C}}

\newcommand{\diff}{\mathop{}\mathopen{}\mathrm{d}}

\DeclareMathOperator{\tr}{tr}
\DeclareMathOperator{\supp}{supp}

\title[Norm Approximation with Linear Growth of Error]{Norm Approximation with Linear Growth of Error for Bosonic Dynamics in the Mean-Field Regime}
\author[Niels Benedikter]{Niels Benedikter}
\address{Università degli Studi di Milano, Dipartimento di Matematica, Via Cesare Saldini 50, 20133 Milano, Italy \& External Scientific Member of Basque Center for Applied Mathematics, Alameda de Mazarredo 14, 48009 Bilbao, Bizkaia, Spain}
\email{\href{mailto:niels.benedikter@unimi.it}{\nolinkurl{niels.benedikter@unimi.it}}}
\author[Chiara Boccato]{Chiara Boccato}
\address{Università di Pisa, Dipartimento di Matematica, Largo Bruno Pontecorvo 5, 56127 Pisa, Italy}
\email{\href{mailto:chiara.boccato@unipi.it}{\nolinkurl{chiara.boccato@unipi.it}}}
\author[Umberto Morellini]{Umberto Morellini}
\address{Università di Pisa, Dipartimento di Matematica, Largo Bruno Pontecorvo 5, 56127 Pisa, Italy}
\email{\href{mailto:umberto.morellini@dm.unipi.it}{\nolinkurl{umberto.morellini@dm.unipi.it}}}
\date{\today}

\begin{document}

\begin{abstract}
    We consider the quantum many-body dynamics of a homogeneous confined Bose gas in three dimensions in the mean-field scaling regime. We study the Cauchy problem for a class of initial data describing finitely many excitations over a Bose–Einstein condensate. Using a rigorous version of Bogolyubov's theory, we prove that the many-body evolution can be approximated in Fock space norm by a quasifree evolution of the excitations with an error which grows linearly in time. In distinction from earlier results, the slow growth in time holds even though the effective equation is not dispersive.
\end{abstract}

\maketitle

\tableofcontents

\section{Introduction}

We consider a system of $N$ bosonic quantum particles in the three-dimensional box $\Lambda=\intcc{-\frac{1}{2}}{\frac{1}{2}}^3$ with periodic boundary conditions. The state of the particles is described by a time-dependent wavefunction $\Psi_N(t)$ in the Hilbert space of totally symmetric square-integrable functions
\begin{equation*}
    L^2_s\left(\Lambda^N\right)=\setst{\psi\in L^2\left(\Lambda^N\right)}{\psi\left(x_{\sigma\left(1\right)},\ldots,x_{\sigma\left(N\right)}\right)=\psi\left(x_1,\ldots,x_N\right) \quad \forall\sigma\in\mathcal{S}_N} \;.
\end{equation*}
The Hamilton operator $H_N$ in the mean-field scaling regime is given by
\begin{equation}\label{eq:HN}
    H_N=\sum_{j=1}^N\left(-\Delta_j\right)+\frac{1}{N}\sum_{i<j}^N V\left(x_i-x_j\right) \;,
\end{equation}
where the Laplacian $-\Delta_j$ describes the kinetic energy of the $j$-th particle and $V$ is the (periodization of) a real-valued, measurable, even function describing the interaction between particles. The mean-field regime is apparent in the choice of the coupling strength as $1/N$, which results in the kinetic energy and the potential energy both having order $N$ in expectation value in typical low-energy states.
We assume that the interaction potential before periodization is in $L^1(\mathbb{R}^3)$ and has Fourier transform that decays sufficiently fast at infinity.
This implies that the Hamiltonian has a self-adjoint realization by the Kato–Rellich theorem.

The quantum dynamics of the system is governed by the Cauchy problem associated to the Schrödinger equation,
\begin{equation}\label{eq:many-particle-schrodinger}
        i\partial_t\Psi_N(t)=H_N\Psi_N(t) \;, \qquad
        \Psi_N\left(0\right)=\Psi_{N,0} \;,
\end{equation}
or, equivalently,
\begin{equation}\label{eq:schrodinger-semigroup}
    \Psi_N(t)=e^{-iH_Nt}\Psi_N\left(0\right) \;.
\end{equation}
The global well-posedness of the Schrödinger equation \eqref{eq:many-particle-schrodinger} is guaranteed by Stone's theorem due to self-adjointness of the Hamiltonian.

To determine the quantum dynamics explicitly, one should solve the Cauchy problem \eqref{eq:many-particle-schrodinger} or, equivalently, compute the strongly continuous unitary group appearing in \eqref{eq:schrodinger-semigroup}. This is an extremely hard task from both a theoretical and a numerical perspective due to the curse of dimensionality: even if the Hilbert space $L^2(\Lambda)$ for a single particle is approximated by a finite-dimensional subspace, the dimension of the $N$-particle space grows rapidly with respect to $N$ (since it is a tensor product). And in experiments with cold atomic gases, already $N$ itself may be of the order of $10^5$. For this reason, a vast literature has been dedicated to rigorously justifying approximations by effective evolution equations. For instance, it is known \cite{CheLeeSch-2011-JSP, ErdYau-2001-arx, KnoPic-2010-CMP, RodSch-2009-CMP} that the time evolution $\Psi_N(t)$ of an initial datum $\Psi_{N,0}$ exhibiting Bose–Einstein condensation in a one-particle state $\varphi\in L^2(\Lambda)$ continues to exhibit Bose–Einstein condensation in an evolved one-particle state $\varphi_t$ given by the solution of the Hartree equation
\begin{equation}\label{eq:non-linear-Hartree-equation}
    i\partial_t\varphi_t=\left(-\Delta+V\ast\abs{\varphi_t}^2\right)\varphi_t \;,
\end{equation}
with initial datum $\varphi_{t=0}=\varphi$. More precisely, the approximation $\Psi_N(t) \approx \varphi_t^{\otimes N}$ holds in trace norm of reduced density matrices. The $k$-particle reduced density matrix $\gamma^{\left(k\right)}_{\Psi_N(t)}$ of $\Psi_N(t)$ is the trace-class operator on the $k$-particle space $L^2_s(\Lambda^{k})$ defined by the integral kernel
\begin{multline*}       \gamma^{\left(k\right)}_{\Psi_N(t)}\left(x_1,\ldots,x_k;y_1,\ldots,y_k\right)\\
    =\int \Psi_N\left(t,x_1,\ldots,x_N\right)\overline{\Psi_N\left(t,y_1,\ldots,y_k,x_{k+1},\ldots,x_N\right)}\diff x_{k+1}\ldots\diff x_N \;.
\end{multline*}
Equivalently, it is the partial trace over $N-k$ tensor factors of the projection operator onto $\Psi_N(t)$.
The mentioned results show that, if $\Psi_{N,0}=\varphi^{\otimes N}$, then the $k$-particle density matrix of $\Psi_N(t)$ converges to that of the Hartree solution $\varphi_t^{\otimes N}$, that is
\begin{equation}\label{eq:density-matrix-convergence}
    \lim_{N\to\infty}\tr_{L^2_s\left(\Lambda^{k}\right)}\abs{\gamma_{\Psi_N(t)}^{\left(k\right)}-\ket{\varphi_t}\bra{\varphi_t}^{\otimes k}}=0 \;,
\end{equation}
for every $k\in\N$ and for every $t\geq 0$. In particular, in \cite{CheLeeSch-2011-JSP}, it is shown that the trace norm in \eqref{eq:density-matrix-convergence} can be bounded by $C e^{Kt}/N$, which is optimal with respect to $N$ (but grows rapidly in time, $C$ and $K$ being positive constants). Moreover, \eqref{eq:density-matrix-convergence} does not provide an approximation for the many-body wavefunction $\Psi_{N}\left(t\right)$ with respect to the norm of  $L^2\left(\Lambda^{N}\right)$. Indeed $\Psi_N(t)$ cannot be expected to stay close to the Hartree state $\varphi_t^{\otimes N}$ in the norm of $L^2\left(\Lambda^{N}\right)$, except in the non-interacting case $V=0$. To obtain a norm approximation in the $N$-particle space, it is not sufficient to approximate the evolution of the condensate. It is instead crucial to take into account the evolution of excitations over the condensate, too.

Different kinds of norm approximations for many-boson dynamics have been derived. In the mean-field regime \cite{Che-2011-ARMA, GriMacMar-2011-AIM, GinVel-1979-CMP,  GriMacMar-2009-CMP, Hep-1974-CMP, LewNamSch-2015-AJM, LewNamSerSol-2014-CPAM, NamRadSof-2025-arx}, it is known that the solution of the many-body Schrödinger equation can be approximated, after removing the condensate wavefunction, by a unitary dynamics on the Fock space of the excitations with a generator quadratic in creation and annihilation operators. This dynamics acts as a family of time-dependent Bogolyubov transformations. Similar norm approximations have been derived for the many-body evolution generated by more singular Hamiltonians of the form
\begin{equation}\label{eq:beta-hamiltonian}
    H_N^\beta= \sum_{j=1}^N\left(-\Delta_j\right)+\frac{1}{N}\sum_{i<j}^N N^{3\beta}V\left(N^\beta\left(x_i-x_j\right)\right),
\end{equation}
first for $\beta\in\intoo{0}{1}$ \cite{BocCenSch-2016-AHP, BreNamNapSch-2019-AIHPC} and, more recently, for the Gross-Pitaevskii regime $\beta=1$ \cite{CarOldSch-2024-AIHPC}. All the cited results (except \cite{Che-2011-ARMA, GriMacMar-2011-AIM}, which concern an initial condensate without excitations, and \cite{NamRadSof-2025-arx}, which we discuss later) have bounds which grow exponentially in time because they rely on Grönwall's lemma. For this reason, those estimates provide norm convergence effectively valid only for rather short times, typically $t=\smallo{\log N}$ coming from error bounds of the form $e^{Kt}/N$.

Similar results have been obtained for many-fermion systems. Because of the Pauli exclusion principle, fermions cannot condense but rather occupy all the low momenta filling the so-called Fermi ball. As a first step, one can study the time evolution of the latter in the mean-field scaling regime and approximations have been established with respect to the trace norm of reduced density matrices \cite{Ben22, BD23, BenPorSch-2014-JMP, BenPorSch-2014-CMP, FroKno-2011-JSP}. In this framework, the Hartree–Fock equation appears in place of the Hartree equation. As in the bosonic case, all estimates exhibit exponential (or even double exponential) growth in time due to Grönwall's lemma. In the fermionic setting, the Fermi ball serves as the counterpart of the bosonic condensate, and accounting for the evolution of the particle-hole excitations around the Fermi ball turns out to be essential for obtaining norm approximations. To the best of our knowledge, the only result addressing this issue is \cite{BenNamPorSch-2021-AHP}, which proves that the many-fermion evolution can be approximated in Fock space norm by a quasifree bosonic dynamics of collective particle-hole modes, relying on a rigorous approximate bosonization technique introduced by \cite{BenNamPorSchSei-2019-CMP, Ben-2020-RMP, BenNamPorSchSei-2021-IM, BenPorSchSei-2023-ARMA}. Some of the ideas of \cite{BenNamPorSch-2021-AHP}, which achieves estimates with only linear growth in time, have served as an important inspiration for the present work.

In the present paper, we derive a norm approximation for the bosonic many-body dynamics in terms of an effective dynamics generated by a  quadratic Hamiltonian. Compared to the existing literature, by considering a class of initial data describing finitely many excitations over a stationary Bose-Einstein condensate, we obtain a bound which grows only linearly in time. In particular, we prove that eigenstates of the effective bosonic quadratic Hamiltonian are almost stationary under the many-body time evolution.

Improving the time-dependence in the bounds of validity of the mean-field approximation has been studied in the whole space $\R^3$, where dispersive estimates permit to improve the typical exponential growth of error estimates into estimates that remain bounded for all times \cite{Die21, DL23, NamRadSof-2025-arx}. One may expect that in this setting, for long times, both the many-body evolution and the effective evolution behave like the free evolution, whose solutions converge weakly to zero, and exploit this fact to show that the many-body evolution remains close to the effective evolution for all times. Our analysis on the torus  complements those results by addressing a setting where dispersive estimates are absent. The mechanism exploited in our work to improve the time-dependence from exponential to linear growth is completely different, and relies only on stationarity of the condensate instead of on dispersivity of the evolution.

\subsection{Second Quantization}
As mentioned, to obtain a norm approximation of the many-body dynamics it is crucial to describe the dynamics of the excitations over the condensate. For this reason, it is convenient to work in second quantization with creation and annihilation operators. The bosonic Fock space over $L^2\left(\Lambda\right)$ is defined as
\begin{equation*}
    \mathscr{F}=\C\oplus\bigoplus_{n\geq 1}L^2_s\left(\Lambda^n\right) \;,
\end{equation*}
and the vacuum vector is denoted by $\Omega=(1,0,0,\ldots)\in\mathscr{F}$. Given $f\in L^2\left(\Lambda\right)$, the creation operator $\check{a}^*\left(f\right)$ is defined as
\begin{equation*}
    \check{a}^*\left(f\right)\left(\zeta_0\oplus\bigoplus_{n\geq 1}\zeta_n\right)=0\ \oplus\ \zeta_0 f\ \oplus\ \bigoplus_{n\geq 1}\left( f\otimes_s \zeta_n \right) \;,
\end{equation*}
and the annihilation operator $\check{a}\left(f\right)$ is its adjoint.

It will be convenient to work in momentum space
\begin{equation*}
\Lambda^*=2\pi\Z^3 \;.
\end{equation*}
Given $p\in\Lambda^*$ and a plane wave $f_p\left(x\right)=e^{-i\scalprod{p}{x}}$,
we define the operators
\begin{equation*}
    a^*_p=\check{a}^*(f_p) \quad \text{and} \quad a_p = \check{a}(f_p)\;,
\end{equation*}
creating and annihilating a particle with momentum $p$, respectively. They satisfy the canonical commutation relations (CCR)
\begin{equation*}
    \left[a_p,a^*_q\right]=\delta_{p,q} \;, \qquad \left[a_p,a_q\right] = 0 = \left[a^*_p,a^*_q\right] \;, \qquad \forall p,q\in\Lambda^* \;.
\end{equation*}
On $\mathscr{F}$, we introduce the number operator $\mathcal{N}$ defined as
\begin{equation*}
    \mathcal{N}=\sum_{p\in\Lambda^*}a^*_p a_p\;.
\end{equation*}
The creation and annihilation operators are unbounded, but relatively bounded with respect to the square root of the number  operator, that is
\begin{equation}\label{eq:N-bound-a-a*}
    \norm{\check{a}(f)\zeta}\leq\norm{f}_{L^2}\norm{\mathcal{N}^{1/2}\zeta}\quad\text{and}\quad\norm{\check{a}^*(f)\zeta}\leq\norm{f}_{L^2}\norm{\left(\mathcal{N}+1\right)^{1/2}\zeta} \;,
\end{equation}
for all $f\in L^2\left(\Lambda\right)$ and $\zeta\in\mathscr{F}$.

The Hamiltonian can be generalized to Fock space as
\begin{equation*}
    H_N=\sum_{p\in\Lambda^*}p^2 a^*_p a_p+\frac{1}{2N}\sum_{p,q,r\in\Lambda^*}\widehat{V}\left(r\right)a^*_{p+r}a^*_qa_p a_{q+r} \;,
\end{equation*}
where $\widehat{V}\left(k\right)=\int_{\mathbb{R}^3}V(x)e^{-ik\cdot x}\diff x$. On the $N$-particle sector of $\mathscr{F}$, this agrees with \eqref{eq:HN}.

If $f_0$ denotes the zero-momentum mode, i.\,e.,
\[
    {f}_0\left(x\right)=1 \quad \text{for all } x\in\Lambda \;,
\]
we define $L^2_\perp\left(\Lambda\right)$ as the orthogonal complement of the one-dimensional space spanned by ${f}_0$ in $L^2\left(\Lambda\right)$. The Fock space of excitations is defined as
\begin{equation*}
    \mathscr{F}_+ = \C\ \oplus\ \bigoplus_{n\geq 1}L^2_\perp\left(\Lambda\right)^{\otimes_s n} \;.
\end{equation*}
The Hilbert space $\mathscr{F}_+$ is generated by creation and annihilation operators $a^*_p$ and $a_p$ with
\begin{equation*}
p\in\Lambda^*_+=2\pi\Z^3\setminus\{0\} \;.
\end{equation*}
On $\mathscr{F}_+$, the number  operator will be denoted by
\begin{equation*}
    \mathcal{N}_+=\sum_{p\in\Lambda^*_+}a^*_p a_p \;.
\end{equation*}
For $N\in\N$, we define the truncated  Fock space of excitations
\begin{equation*}
    \mathscr{F}^{\leq N}_+ = \C\ \oplus\ \bigoplus_{n=1}^NL^2_\perp\left(\Lambda\right)^{\otimes_s n} \;.
\end{equation*}
Finally, on $\mathscr{F}^{\leq N}_+$, we introduce modified annihilation and creation operators, setting
\begin{equation}\label{eq:bb*-operators-def}
    b_p=\sqrt{\frac{N-\mathcal{N}_+}{N}}a_p \;, \quad b^*_p=a^*_p\sqrt{\frac{N-\mathcal{N}_+}{N}} \;, \quad \text{for } p\in\Lambda^*_+ \;.
\end{equation}
Note that these operators leave $\mathscr{F}^{\leq N}_+$ invariant, in contrast to the standard creation and annihilation operators. They satisfy the commutation relations
\begin{equation}
    \begin{gathered}\label{eq:bbstarCom}
        \left[b_p,b^*_q\right]=\left(1-\frac{\mathcal{N}_+}{N}\right)\delta_{p,q}-\frac{1}{N}a^*_q a_p=\delta_{p,q}+\mathcal{E}\left(p,q\right)\;, \qquad
    \left[b_p,b_q\right]=0=\left[b^*_p,b^*_q\right] \;.
    \end{gathered}
\end{equation}
The modified creation and annihilation operators are almost bosonic, meaning that the error term $\mathcal{E}\left(p,q\right)$ converges to zero in the limit of $N \to \infty$; more precisely, the operator $\mathcal{E}\left(p,q\right)$, for any $p,q\in\Lambda^*_+$, commutes with the number  operator $\mathcal{N}_+$ on $\mathscr{F}_+$ and, as a consequence of \eqref{eq:N-bound-a-a*}, satisfies
    \begin{equation}    \label{eq:quasibos}
        \norm{\mathcal{E}\left(p,q\right)\zeta}\leq C N^{-1}\norm{\mathcal{N}_+ \zeta}\;, \quad \text{for all } \zeta\in\mathscr{F}_+ \;.
    \end{equation}
Moreover, the following commutation relations will be useful later:
\begin{equation}\label{eq:commutation-rules}
    \left[b_p,a^*_q a_r\right]=\delta_{p,q}b_r\quad\text{and}\quad\left[b^*_p,a^*_q a_r\right]=-\delta_{p,r}b^*_q \;,
\end{equation}
Furthermore,
\begin{equation}\label{eq:Nb}
 \left[b_p,\mathcal{N}_+\right]=b_p\qquad \text{and} \qquad\left[b^*_p,\mathcal{N}_+\right]=-b^*_p\;.
\end{equation}
Finally, we know that
\begin{equation}\label{eq:b-b*-bounds}
    \norm{b_p\zeta}\leq\norm{\mathcal{N}_+^{1/2}\zeta}\quad\text{and}\quad\norm{b^*_p\zeta}\leq\norm{\left(\mathcal{N}_++1\right)^{1/2}\zeta}\;,
\end{equation}
for all $\zeta\in\mathscr{F}^{\leq N}_+$. In particular $b_p$ and $b^*_p$ are bounded operators, since $\mathcal{N}_+\leq N$ on $\mathscr{F}^{\leq N}_+$.

\subsection{Excitation Hamiltonian}
It is convenient to separate excited particles from the particles in the condensate by means of the unitary map
\begin{equation}\label{eq:unitary-map}
    \begin{aligned}
        U_N:L^2_s(\Lambda^N)&\to\mathscr{F}^{\leq N}_+\\
        \Psi\quad&\mapsto\psi_0\oplus\psi_1\oplus\ldots\oplus\psi_N \;,
    \end{aligned}
\end{equation}
where $\psi_n\in (L^2_\perp(\Lambda))^{\otimes_s n}$, $0\leq n\leq N$, is defined by the unique decomposition of $\Psi\in L^2_s\left(\Lambda^N\right)$ as
\begin{equation*}
    \Psi=\psi_0 f_0^{\otimes N}+\psi_1\otimes_s f_0^{\otimes\left(N-1\right)}+\psi_2\otimes_s f_0^{\otimes\left(N-2\right)}+\ldots+\psi_N \;.
\end{equation*}
This unitary operator $U_N$ was first introduced in \cite{LewNamSerSol-2014-CPAM}. Observe that
\begin{equation}
    \mathscr{F}^{\leq N}_+\subset \mathscr{F}_+\subset\mathscr{F}\;,
\end{equation}
and therefore we can always consider $U_N$ as a partial isometry from $L^2_s\left(\Lambda^N\right)$ to $\mathscr{F}$, where $U_N^*$ is extended by $0$ outside of $\mathscr{F}_+^{\leq N}$.
Conjugating creation and annihilation operators by $U_N$ (note that products $a^*_pa_q$ leave the Fock space sector $L^2_s\left(\Lambda^N\right)$ invariant), we find that, for any $p,q\in\Lambda^*_+$,
\begin{align*}
    U_N a^*_0a_0U_N^*&=N-\mathcal{N}_+ \;, &
    U_N a^*_pa_0U_N^*&=a^*_p\sqrt{N-\mathcal{N}_+} \;,\\
    U_N a^*_0a_pU_N^*&=\sqrt{N-\mathcal{N}_+}a_p \;, &
    U_N a^*_pa_qU_N^*&=a^*_pa_q \;.
\end{align*}
In particular, for any $p\in\Lambda^*_+$, we have
\begin{equation*}
    U_Na^*_p\frac{a_0}{\sqrt{N}}U_N^*=b^*_p \;,
\end{equation*}
which shows that the modified creation operator excites a particle from the condensate. Similarly, the modified annihilation operator annihilates an excitation, moving a particle back to the condensate. Compared to the standard creation and annihilation operators $a^*$ and $a$, the modified operators $b^*$ and $b$ create or annihilate excitations preserving at the same time the total number of particles.

The excitation Hamiltonian is defined as
\begin{equation}\label{eq:LN}
\mathcal{L}_N=U_NH_N U^*_N:\mathscr{F}^{\leq N}_+\to\mathscr{F}^{\leq N}_+ \;,
\end{equation}
and has the explicit form (see \cite[Equation~(3.4)]{BocBreCenSch-2019-AM} or \cite[Equation~(3.3)]{BocBreCenSch-2020-JEMS})
\begin{equation}\label{eq:excitations-hamiltonian-decomposition}
    \mathcal{L}_N=\mathcal{L}_N^{\left(0\right)}+\mathcal{L}_N^{\left(2\right)}+\mathcal{L}_N^{\left(3\right)}+\mathcal{L}_N^{\left(4\right)} \;,
\end{equation}
where
\begin{equation}
\begin{split}
\label{eq:LNterms}
    \mathcal{L}_N^{\left(0\right)}&=\frac{N-1}{2N}\widehat{V}\left(0\right)\left(N-\mathcal{N}_+\right)+\frac{\widehat{V}\left(0\right)}{2N}\mathcal{N}_+\left(N-\mathcal{N}_+\right) \;,\\
    \mathcal{L}_N^{\left(2\right)}&=\sum_{p\in\Lambda^*_+}p^2 a^*_pa_p+\sum_{p\in\Lambda^*_+}\widehat{V}\left(p\right)\left[b^*_pb_p-\frac{1}{N}a^*_pa_p\right]+\frac{1}{2}\sum_{p\in\Lambda^*_+}\widehat{V}\left(p\right)\left[b^*_pb^*_{-p}+b_p b_{-p}\right] \;,\\
    \mathcal{L}_N^{\left(3\right)}&=\frac{1}{\sqrt{N}}\sum_{\substack{p,q\in\Lambda^*_+\\p+q\neq 0}}\widehat{V}\left(p\right)\left[b^*_{p+q}a^*_{-p}a_q+a^*_qa_{-p}b_{p+q}\right] \;,\\
    \mathcal{L}_N^{\left(4\right)}&=\frac{1}{2N}\sum_{\substack{p,q\in\Lambda^*_+\\r\in\Lambda^*\\r\neq-p,-q}}\widehat{V}\left(r\right)a^*_{p+r}a^*_qa_pa_{q+r} \;.
\end{split}
\end{equation}

The transformation $U_N$ extracts the contribution of the condensate. To describe the dynamics of excitations we need to conjugate the Hamiltonian by a Bogolyubov transformation, defined as follows. We introduce $\eta=\left(\eta_p\right)_{p\in\Lambda^*_+}\subset\R$ such that $\eta_p=0$ for $p\in\, P_L^c$, and
\begin{equation}\label{eq:etaMF}
    \tanh\left(2\eta_p\right)=-\frac{\widehat{V}\left(p\right)}{p^2+\widehat{V}\left(p\right)} \;, \qquad \text{for } p\in\, P_L\;,
\end{equation}
where
\begin{equation}\label{eq:PL}
    P_L=\{p\in\Lambda^*_+:|p|\leq N^\alpha\}\;,
\end{equation}
with $0<\alpha<1/5$ a parameter to be optimized. This is equivalent to
\begin{equation*}
    \eta_p = \begin{cases}\frac{1}{4}\log \left(\frac{p^2}{p^2+2\widehat{V}\left(p\right)}\right)=\frac{1}{4}\log\left(1-\frac{2\widehat{V}\left(p\right)}{p^2+2\widehat{V}\left(p\right)}\right)  \qquad \text{for } p\in\, P_L\;, \\
                0  \qquad \text{for } p\in\, P_L^c \;.
             \end{cases}
\end{equation*}
Note that, for some constant $C_V > 0$, we have
\begin{equation}\label{eq:bogolyubov-kernel-bound}
    \abs{\eta_p}\leq C_V p^{-2} \abs{\widehat{V}(p)}
\end{equation}
and
\begin{equation}\label{eq:etaL1}
    \|\eta\|_{\ell^1}\leq C_V \;.
\end{equation}
Using these coefficients, we define the generalized Bogolyubov transformation
\begin{equation}\label{eq:bogolyubov-diagonalisation}
    T=e^{B\left(\eta\right)}:\mathscr{F}^{\leq N}_+\to\mathscr{F}^{\leq N}_+
\end{equation}
using
\begin{equation}\label{eq:bogolyubov-kernel}
    B\left(\eta\right)=\frac{1}{2}\sum_{p\in P_L}\eta_p\left(b^*_pb^*_{-p}-b_pb_{-p}\right) \;,
\end{equation}
To be precise, $T$ would implement a Bogolyubov transformation on the bosonic Fock space if the $b$- and $b^*$-operators satisfied canonical commutation relations without the error term $\mathcal{E}(p,q)$. As given, $T$ is a unitary transformation on Fock space which only in the limit $N \to \infty$, and applied to suitable states, can be expected to approximate a Bogolyubov transformation; therefore we call it a generalized Bogolyubov transformation. It acts on the $b_p$- and $b_p^*$-operators, for $p\in\, P_L$, as
\begin{equation}\label{eq:genbT}
    T^*b_pT=\cosh(\eta_p)b_p+\sinh(\eta_p)b^*_{-p}+D_p\;,
\end{equation}
where $D_p$ will be explicitly defined in \eqref{eq:Dp-def} and is small on states with few excitations.
With the choice \eqref{eq:etaMF}, $T$ diagonalizes (in the sense of Bogolyubov theory) the quadratic part $\mathcal{L}^{(2)}_N$ of the excitation Hamiltonian, restricted additionally to only momenta in the set $P_L$, as defined in \eqref{eq:PL}. This restriction is crucial to control the error terms coming from the diagonalization; modes corresponding to higher momenta are not diagonalized, but they don't create too many excitations due to the fast decay of $\widehat V$.

\medskip

We introduce a new excitation Hamiltonian $\mathcal{M}_N:\mathscr{F}^{\leq N}_+\to\mathscr{F}^{\leq N}_+$ by setting
\begin{align}\label{eq:MN}
    \mathcal{M}_N&=T^*\mathcal{L}_N T=T^*U_NH_N U_N^*T\;.
\end{align}
In the next section we are going to prove that, for suitable initial states,  the dynamics of excitations generated by $\mathcal{M}_N$ can be approximated by the dynamics generated by the quadratic Hamiltonian
\begin{equation}\label{eq:diagonal-bogolyubov-hamiltonian}
    \mathbb{H}_{\textnormal{B}}=\sum_{p\in P_L}\sqrt{p^4+2\widehat{V}\left(p\right)p^2}a^*_pa_p\;,
\end{equation}
with an overall phase given by $e^{-iE_Nt}$, with
\begin{equation}\label{eq:phase}
    E_N=\frac{N}{2}\widehat{V}\left(0\right)-\frac{1}{2}\sum_{p\in \Lambda^*}\left(p^2+\widehat{V}\left(p\right)-\sqrt{p^4+2p^2\widehat{V}\left(p\right)}\right)\;.
\end{equation}
Notice that \eqref{eq:diagonal-bogolyubov-hamiltonian} and \eqref{eq:phase} are the mean-field excitation spectrum and the mean-field energy, respectively, in \cite[Theorem~1]{Sei-2011-CMP} (note however that we consider slightly more general interaction potentials).

\subsection{Main Result}
We study the time evolution of initial data describing finitely many excitations over a Bose–Einstein condensate. Let $U_N$ be the unitary map defined in \eqref{eq:unitary-map} and $T$ the generalized diagonalizing Bogolyubov transformation defined in \eqref{eq:bogolyubov-diagonalisation}. We consider the many-body Schrödinger equation \eqref{eq:many-particle-schrodinger} with the initial state
\begin{equation}\label{eq:initial-state}
    \Psi_{N,0}=U^*_NT\xi\in L^2_s({\Lambda^N})\;, \quad \xi=\frac{1}{Z_m}b^*\left(\varphi_1\right)\ldots b^*\left(\varphi_m\right)\Omega \;,
\end{equation}
where
\begin{equation}\label{eq:b-star-decomposition}
    b^*\left(\varphi_j\right)=\sum_{q\in\Lambda^*_+}\varphi_j\left(q\right)b^*_q \;,
\end{equation}
and $\varphi_1,\ldots,\varphi_m\in \ell^2(\Lambda^*_+)$ are normalized one-boson wavefunctions such that $\supp\varphi_j\subset B\left(0,\sqrt{\sigma}\right)$ for some $\sigma>1$. The normalization constant $Z_m$ is chosen such that $\norm{\xi}=1$. Moreover we define the family of time-dependent states
\begin{equation}\label{eq:evolved-initial-state}
    \xi_t=\frac{1}{Z_m}b^*\left(\varphi_{1,t}\right)\ldots b^*\left(\varphi_{m,t}\right)\Omega\;, \quad t\in\R \;,
\end{equation}
where
\begin{equation}\label{eq:dynExc}
    \varphi_{j,t}=e^{-iH_Bt}\varphi_j \;,\quad j=1,\ldots,m \;,
\end{equation}
with the operator $H_B$ being the first-quantized version of \eqref{eq:diagonal-bogolyubov-hamiltonian}, in Fourier space represented as
\begin{equation*}
    H_B(p)=\sqrt{p^4+2\widehat{V}\left(p\right)p^2}\;.
\end{equation*}
Since $H_B$ is a multiplication operator, the support condition $\supp\varphi_{j,t}\subset B\left(0,\sqrt{\sigma}\right)$ is preserved for all times $t$. The state $\xi_t$ can be viewed as an approximate $m$-particle bosonic state, where every $\varphi_j$ evolves according to a one-body quantum dynamics generated by the Hamiltonian $H_B$. The vector $\xi_t$ is normalized too; this is proven in Lemma~\ref{stat:xit-normalized} below.

\medskip

The next theorem is our main result. It provides a norm approximation for the $N$-body evolution of $\psi=U^*_NT\xi$ in terms of the explicit state $U^*_NT\xi_t$ when $N\xrightarrow{}+\infty$.
\begin{thm}[Main result] \label{stat:main-thm}
Assume that $V\in L^1\left(\mathbb{R}^3\right)$ is spherically symmetric and its Fourier transform satisfies $p^2+2\widehat V(p)>0$ for every $p\in\Lambda_+^*$. Assume furthermore that there exists $p_0>0$ such that for $|p|\geq p_0$ we have
 $|\widehat{V}(p)|\leq C/|p|^{3+\varepsilon}$ with some $\varepsilon>0$. Let $\xi$ be as in \eqref{eq:initial-state} and $\xi_t$ be as in \eqref{eq:evolved-initial-state}. Then, there exists a constant $C_{m,V,\sigma}>0$, depending only on $m$, $V$ and $\sigma$, such that for all $t \in \R$ we have, for $N$ large enough,
    \begin{equation}\label{eq:norm-approximation}
        \norm{e^{-iH_Nt}U^*_NT\xi-e^{-iE_Nt}U^*_NT\xi_t}\leq\begin{cases}
  C_{m,V,\sigma}N^{-1/2}\abs{t}\;,\quad\text{for }\varepsilon\geq5 \;,\\
    C_{m,V,\sigma}\varepsilon^{-1}N^{-\frac{\varepsilon}{5+\varepsilon}}\abs{t}\;,\quad\text{for }\varepsilon<5 \;.
\end{cases}
    \end{equation}
\end{thm}

\begin{rmk}
    \begin{itemize}
        \item[(i)] One can generalize the theorem to initial data in which the number $m$ of excitations grows slowly as $N$ goes to infinity, as long as $m^2\ll N^{1/2}$ for $\varepsilon\geq5$, or $m^{2}\ll N^{\varepsilon/(5+\varepsilon)}$ for $0 < \varepsilon<5$. Indeed  $C_{m,V,\sigma}\leq m^{2}C_{V,\sigma}$  where $C_{V,\sigma}$ depends only on  $V$ and $\sigma$, as one can see from \eqref{eq:cEN}.
        \item[(ii)] Let $\varphi$ be an eigenvector of $H_B$, such that $H_B\varphi=\lambda\varphi$ with eigenvalue $\lambda>0$. For $\xi=Z_1^{-1}b^*(\varphi)\Omega$, we have $\xi_t=Z_1^{-1}b^*(e^{-iH_Bt}\varphi)\Omega = e^{-i\lambda t}\xi$. Then,
        \[
        \norm{e^{-iH_Nt}U^*_NT\xi-e^{-i\left(E_N+\lambda\right)t}U^*_NT\xi}\leq C_{m,V,\sigma}N^{-\gamma}\abs{t}\;, \quad \text{for any }t\in\R \;,
        \]
        for a suitable $\gamma>0$ depending on $\varepsilon$.
        Therefore $U^*_NT\xi$ is an almost stationary state of the many-body Hamiltonian $H_N$ up to times of order $o(N^\gamma)$. Thus, \cref{stat:main-thm} provides a method for the explicit construction of approximate eigenstates of the many-body Hamiltonian.
        \item[(iii)] By \cite{Sei-2011-CMP} and \cite[Theorem~1.1]{BocBreCenSch-2019-AM}, we know that the equality between the spectrum of $H_N-E_N$ and the one of $\mathbb{H}_{\textnormal{B}}$ in \eqref{eq:diagonal-bogolyubov-hamiltonian} holds below a certain threshold $\sigma$, up to an error small in $N$. For this reason, we require $\supp\varphi_j\subset B\left(0,\sqrt{\sigma}\right)$.
        The analysis in \cref{sec:error-norm-estimates} and \cref{sec:proof-of-prop} shows that $C_{m,V,\sigma}$ grows as $\sigma^{5/2}$.
        \item[(iv)] The decay assumption on $\widehat{V}$ is necessary to control the errors from the diagonalization of the effective quadratic Hamiltonian (see Proposition \ref{prop:errorMN}). This was not necessary in previous results providing norm approximations for the dynamics, such as \cite{LewNamSch-2015-AJM}, whose effective Hamiltonian was not diagonal. In the fermionic setting, the norm approximation with linear behavior in time for the error in \cite{BenNamPorSch-2021-AHP} requires a compactly supported $\widehat V$.
        \item[(v)] For $\widehat{V}$ decaying fast enough, the convergence in \eqref{eq:norm-approximation} attains the optimal $N$-dependence predicted in \cite[Remark~3]{LewNamSch-2015-AJM}.
    \end{itemize}
\end{rmk}

\paragraph{\textbf{Organization of the paper}} In \cref{sec:reduction-of-norm-approximation}, we show how \cref{stat:main-thm} reduces to finding uniform-in-time estimates of the error arising from the excitation Hamiltonian $\cM_N$ in \eqref{eq:MN}. In \cref{sec:error-norm-estimates}, we provide norm estimates of the error terms in the excitation Hamiltonian $\mathcal{L}_N=U_N H_N U_N^*$. We then prove that these estimates are stable under conjugation by $T$. In \cref{sec:proof-of-prop}, we control errors resulting from the diagonal quadratic approximation of $\cM_N$. The general strategy is inspired by \cite{BenNamPorSch-2021-AHP}.

\section{Reduction of Norm Approximation to Error Norm Estimates}\label{sec:reduction-of-norm-approximation}
To prove Theorem \ref{stat:main-thm},  we will show that, when restricting to states $\xi_t$ of the form \eqref{eq:evolved-initial-state}, the excitation Hamiltonian $\cM_N$ can be approximated by the diagonal quadratic Hamiltonian $\mathbb{H}_{\textnormal{B}}$ in \eqref{eq:diagonal-bogolyubov-hamiltonian}, up to a small error whose norm does not depend on $t$, for any $t\in \mathbb{R}$. This is the content of the following proposition, whose proof is postponed to Section \ref{sec:proof-of-prop}. Moreover, we will show that $\xi_t$ remains normalized. These results lead to  Theorem \ref{stat:main-thm}, which we prove at the end of the section.

\begin{prop}\label{prop:errorMN}
Let $\mathcal{M}_N$ be defined as in \eqref{eq:MN}, $E_N$ as in \eqref{eq:phase}, $\mathbb{H}_{\textnormal{B}}$ as in
\eqref{eq:diagonal-bogolyubov-hamiltonian} and $\xi_t$ as in \eqref{eq:evolved-initial-state}. Then we have
  \begin{align}\label{eq:cM}
    \mathcal{M}_N&=E_N+\mathbb{H}_{\textnormal{B}}+\mathcal{E}_N \;,
\end{align}
with the error term $\mathcal{E}_N$  satisfying, for $N$ large enough and for any $t\in\R$,
\begin{equation}\label{eq:cEN}
\|\mathcal{E}_N\xi_t\|\leq  C_{V,\sigma}m^{3/2}N^{-1/2}+C_V\varepsilon^{-1}m N^{-\varepsilon\alpha}+C_Vm^2N^{5\alpha-1}\;.
\end{equation}
for constants $C_{V,\sigma}$ and $C_V$, depending on $V$ and $\sigma$, and only on $V$, respectively. The parameter $\alpha$ is the one chosen in \eqref{eq:PL} and may be optimized eventually.
\end{prop}

Next, we prove that the constant $Z_m$, defined such that $\xi$ in \eqref{eq:initial-state} is normalized, makes also the time-evolved state $\xi_t$ in \eqref{eq:evolved-initial-state} normalized.
\begin{lem}\label{stat:xit-normalized}
%     Let $\xi_t$ be defined in \eqref{eq:evolved-initial-state}.
    For any $t\in\R$, we have $\|\xi_t\|=1$.
\end{lem}
\begin{proof}
    Define
    \begin{equation*}
        c_{N,m}^2=\prod_{j=0}^{m-1}\left(1-\frac{j}{N}\right)\;.
    \end{equation*}
    On the one hand, one can easily check that
    \begin{equation*}
        b^*(\varphi_{1,t})\ldots b^*(\varphi_{m,t})\Omega=c_{N,m}a^*(\varphi_{1,t})\ldots a^*(\varphi_{m,t})\Omega\;.
    \end{equation*}
    Indeed, the rightmost creation operator acts on the vacuum, the next on the one-particle sector, and so on. Each of these $b^*$-operators generates a scalar factor, and their product yields exactly $c_{N,m}$. On the other hand,
    \begin{align*}
        &\scalprod{a^*\left(\varphi_{1,t}\right)\ldots a^*\left(\varphi_{m,t}\right)\Omega}{a^*\left(\varphi_{1,t}\right)\ldots a^*\left(\varphi_{m,t}\right)\Omega}\\
        &\hspace{4cm}=\scalprod{a\left(\varphi_{1,t}\right)a^*\left(\varphi_{1,t}\right)\ldots a^*\left(\varphi_{m,t}\right)\Omega}{a^*\left(\varphi_{2,t}\right)\ldots a^*\left(\varphi_{m,t}\right)\Omega}\\
        &\hspace{4cm}=\scalprod{\left[a\left(\varphi_{1,t}\right),a^*\left(\varphi_{1,t}\right)\ldots a^*\left(\varphi_{m,t}\right)\right]\Omega}{a^*\left(\varphi_{2,t}\right)\ldots a^*\left(\varphi_{m,t}\right)\Omega} \;.
    \end{align*}
    Now, the commutator equals
    \begin{multline*}
        \left[a\left(\varphi_{1,t}\right),a^*\left(\varphi_{1,t}\right)\ldots a^*\left(\varphi_{m,t}\right)\right]\\
        =\sum_{j=1}^m a^*\left(\varphi_{1,t}\right)\ldots a^*\left(\varphi_{j-1,t}\right)\left[a\left(\varphi_{1,t}\right),a^*\left(\varphi_{j,t}\right)\right]a^*\left(\varphi_{j+1,t}\right)\ldots a^*\left(\varphi_{m,t}\right) \;.
    \end{multline*}
    Hence,
    \begin{multline*}
        \scalprod{a^*\left(\varphi_{1,t}\right)\ldots a^*\left(\varphi_{m,t}\right)\Omega}{a^*\left(\varphi_{1,t}\right)\ldots a^*\left(\varphi_{m,t}\right)\Omega}\\
        =\sum_{j=1}^m\scalprod{\varphi_{j,t}}{\varphi_{1,t}}\scalprod{a^*\left(\varphi_{1,t}\right)\ldots a^*\left(\varphi_{j-1,t}\right)a^*\left(\varphi_{j+1,t}\right)\ldots a^*\left(\varphi_{m,t}\right)\Omega}{a^*\left(\varphi_{2,t}\right)\ldots a^*\left(\varphi_{m,t}\right)\Omega}\;.
    \end{multline*}
    Iterating these steps $m$ times, we get
    \begin{equation*}
        \scalprod{b^*\left(\varphi_{1,t}\right)\ldots b^*\left(\varphi_{m,t}\right)\Omega}{b^*\left(\varphi_{1,t}\right)\ldots b^*\left(\varphi_{m,t}\right)\Omega}=c_{N,m}^2\sum_{\sigma\in\mathcal{S}_m}\prod_{j=1}^m\scalprod{\varphi_{j,t}}{\varphi_{\sigma\left(j\right),t}}\;.
    \end{equation*}
    Notice that the left-hand side equals $Z_m^2$ at $t=0$. Moreover, the scalar products on the right-hand side are independent of $t$, by unitarity of the dynamics generated by $H_B$. This concludes the proof.
\end{proof}

We now prove \cref{stat:main-thm} by estimating the difference in \eqref{eq:norm-approximation} by uniform-in-time norm estimates of $\mathcal{E}_N$ (compare to \cite[Lemma 3.1]{BenNamPorSch-2021-AHP}).
\begin{proof}[Proof of Theorem \ref{stat:main-thm}] Let $\xi$ be as in \eqref{eq:initial-state}, $\xi_t$ be as in \eqref{eq:evolved-initial-state} and $\mathcal{E}_N$ as in \eqref{eq:EN-error}.
    Using the unitarity of $e^{-iH_Nt}$, $U_N$, and $T$, and Proposition \ref{prop:errorMN}, we rewrite, for $t>0$,
    \begin{align*}
        & \norm{e^{-iH_Nt}U^*_NT\xi-e^{-iE_Nt}U^*_NT\xi_t} =\norm{U_N^*T\xi-e^{i\left(H_N-E_N\right)t}U_N^*T\xi_t}\\
        &=\norm{\int_0^t\frac{\diff}{\diff s}\left(e^{i\left(H_N-E_N\right)s}U_N^*T\xi_s\right)\diff s}
         \leq\int_0^t\norm{T^*U_N\left(H_N-E_N\right)U_N^*T\xi_s-i\partial_s\xi_s}\diff s\\
        &=\int_0^t \norm{\left(\mathbb{H}_{\textnormal{B}}+\mathcal{E}_N\right)\xi_s-i\partial_s\xi_s}\diff s \;.
    \end{align*}
On the one hand, using \eqref{eq:dynExc},  we have
    \begin{align*}
        i\partial_s\xi_s&=\frac{1}{Z_m}\sum_{j=1}^m b^*\left(\varphi_{1,s}\right)\ldots b^*\left(i\partial_s\varphi_{j,s}\right)\ldots b^*\left(\varphi_{m,s}\right)\Omega\\
        &=\frac{1}{Z_m}\sum_{j=1}^m b^*\left(\varphi_{1,s}\right)\ldots b^*\left(H_B\varphi_{j,s}\right)\ldots b^*\left(\varphi_{m,s}\right)\Omega \;.
    \end{align*}
On the other hand, using that $\mathbb{H}_{\textnormal{B}}\Omega=0$, we get
    \begin{align*}
        \mathbb{H}_{\textnormal{B}}\xi_s&=\frac{1}{Z_m}\mathbb{H}_{\textnormal{B}} b^*\left(\varphi_{1,s}\right)\ldots b^*\left(\varphi_{m,s}\right)\Omega=\frac{1}{Z_m}\left[\mathbb{H}_{\textnormal{B}},b^*\left(\varphi_{1,s}\right)\ldots b^*\left(\varphi_{m,s}\right)\right]\Omega\\
        &=\frac{1}{Z_m}\sum_{j=1}^m b^*\left(\varphi_{1,s}\right)\ldots b^*\left(\varphi_{j-1,s}\right)\left[\mathbb{H}_{\textnormal{B}},b^*\left(\varphi_{j,s}\right)\right]b^*\left(\varphi_{j+1,s}\right)\ldots b^*\left(\varphi_{m,s}\right)\Omega \;.
    \end{align*}
        Recalling the definition \eqref{eq:diagonal-bogolyubov-hamiltonian} of $\mathbb{H}_{\textnormal{B}}$,
     using \eqref{eq:b-star-decomposition}, and \eqref{eq:commutation-rules}, we get
    \begin{align*}
        \left[a^*_pa_p,b^*\left(\varphi_{j,s}\right)\right] & =\sum_{q\in\Lambda^*_+\cap B\left(0,\sqrt{\sigma}\right)}\varphi_{j,s}\left(q\right)\left[a^*_pa_p,b^*_q\right]
        =\varphi_{j,s}\left(p\right)b^*_p \;,
    \end{align*}
    which then implies
    \begin{equation*}
        \left[\mathbb{H}_{\textnormal{B}},b^*\left(\varphi_{j,s}\right)\right]=\sum_{p\in\Lambda^*_+\cap B(0,\sqrt{\sigma})}\sqrt{p^4+2\widehat{V}\left(p\right)p^2}\varphi_{j,s}\left(p\right)b^*_p=b^*\left(H_B\varphi_{j,s}\right) \;.
    \end{equation*}
    Thus, we have the cancellation $\mathbb{H}_{\textnormal{B}}\xi_s - i\partial_s\xi_s = 0$, which means that, for any $t\in\R$,
    \begin{align*}
        & \norm{e^{-iH_Nt}U^*_NT\xi-e^{-iE_Nt}U^*_NT\xi_t} \leq\int_0^t\norm{\mathcal{E}_N\xi_s}\diff s\;.
    \end{align*}
We use now the bound \eqref{eq:cEN} in Proposition~\ref{prop:errorMN}. Optimizing over the choice of $\alpha$,  for a fixed $\varepsilon$, gives \eqref{eq:norm-approximation}.
\end{proof}

\section{Error Norm Estimates and Stability under the Generalized Bogolyubov Transformation}
\label{sec:stability-of-error-norm-estimates}
In this section, we analyze the excitation Hamiltonian $\mathcal{L}_N$ in \eqref{eq:LN} and prove that, on states of the form \eqref{eq:evolved-initial-state}, its main contribution is quadratic in the $b$- and $b^*$-operators, up to an error that can be bounded uniformly in $t$ (see Proposition \ref{prop:error1} below). Moreover,
the uniform-in-time error norm estimates  are stable under conjugation by $T$.
A first important observation in this direction is the following lemma, whose proof can be found in \cite[Lemma~3.1]{BrSchBrSch} (a similar result has been previously established in \cite{Sei-2011-CMP}).
\begin{lem}\label{lm:Ngrow}
For every $n \in \mathbb{N}$, there exists a constant $C > 0$ such that, on $\cF$,
\begin{equation}\label{eq:bd-Beta} e^{-B(\eta)} (\cN_+ +1)^{n} e^{B(\eta)} \leq C e^{C \| \eta \|} (\cN_+ +1)^{n}\;,
\end{equation}
for all $\eta \in \ell^2 (\Lambda^*)$.
\end{lem}

Lemma~\ref{lm:Ngrow} above will be used to prove the following proposition.

\label{sec:error-norm-estimates}
\begin{prop}\label{prop:error1}
Let $\mathcal{L}_N$ be defined as in \eqref{eq:LN}. Then, we have
\begin{align}\label{eq:lN}
\mathcal{L}_N&=\frac{N-1}{2}\widehat{V}(0)+\mathcal{Q}_N+\mathcal{E}_{\mathcal{L}_N}\;,
\end{align}
with
\begin{align}\label{eq:QPL}
\mathcal{Q}_N=    \sum_{p\in P_L}\Big[\big(p^2+\widehat{V}\left(p\right)\big)b^*_pb_p+\frac{1}{2}\widehat{V}\left(p\right)\left(b^*_pb^*_{-p}+b_pb_{-p}\right)\Big] \;,
\end{align}
where $P_L$ has been defined in \eqref{eq:PL},  and
\begin{align}\label{eq:ELN}
\mathcal{E}_{\mathcal{L}_N}&=\widetilde{\mathcal{L}}_N^{\left(0\right)}+\mathbb{Q}_2^{\left(1\right)}+\mathbb{Q}_2^{\left(2\right)}+\mathbb{Q}_2^{\left(3\right)}+ \mathcal{L}_N^{\left(3\right)}+\mathcal{L}_N^{\left(4\right)}\;,
\end{align}

with
\begin{equation}\begin{split}\label{eq:errorlist1}
\widetilde{\mathcal{L}}_N^{\left(0\right)}&=\frac{\widehat{V}\left(0\right)}{2N}\left(\mathcal{N}_+-\mathcal{N}_+^2\right)\\
    \mathbb{Q}_2^{\left(1\right)} &=
    \frac{1}{N}\sum_{p\in\Lambda^*_+}p^2 a^*_pa_p(\mathcal{N}_+-1)\;, \\
    \mathbb{Q}_2^{\left(2\right)} & = -\frac{1}{N}\sum_{p\in\Lambda^*_+}\widehat{V}\left(p\right)a^*_pa_p\;,\\
    \mathbb{Q}_2^{\left(3\right)}& =    \sum_{p\in P_L^c\backslash \{0\}}\Big[\big(p^2+\widehat{V}\left(p\right)\big)b^*_pb_p+\frac{1}{2}\widehat{V}\left(p\right)\left(b^*_pb^*_{-p}+b_pb_{-p}\right)\Big]\;,
\end{split}
\end{equation}
and $\mathcal{L}_N^{\left(3\right)}$ and $\mathcal{L}_N^{\left(4\right)}$ defined as in \eqref{eq:LNterms}.
Recalling the definition for $T$ in~\eqref{eq:bogolyubov-diagonalisation}, the error $\mathcal{E}_{\mathcal{L}_N}$ satisfies, for $N$ large enough and for any $t\in\R$,
\begin{equation}\label{eq:estELN}
\norm{\mathcal{E}_{\mathcal{L}_N}\xi_t},\norm{\cE_{\cL_N}T\xi_t} \leq C_{V,\sigma}m^{3/2} N^{-1/2}+C_V\varepsilon^{-1}m N^{-\varepsilon\alpha} \;,
\end{equation}
for constants $C_{V,\sigma}$ and $C_V$, depending on $V$ and $\sigma$, and only on $V$, respectively. The parameter $\alpha$ is the one chosen in \eqref{eq:PL} and may be optimized eventually.
\end{prop}

\begin{proof}[Proof of Proposition \ref{prop:error1}] Using the expansion \eqref{eq:LNterms} for $\cL_N$, a calculation shows that \eqref{eq:lN} holds with the remainder $\cE_{\cL_N}$ as in \eqref{eq:ELN}.
Using \eqref{eq:Nb}, we have, for any $\zeta \in \mathscr{F}^{\leq N}_+$,
\begin{align}\label{eq:er1}
\|\widetilde{\mathcal{L}}_N^{\left(0\right)}\zeta\|&= \frac{|\widehat{V}\left(0\right)|}{2N}\|\left(\mathcal{N}_+-\mathcal{N}_+^2\right)\zeta\|\;,
\end{align}
which means that
\begin{align}\label{eq:er11}
\|\widetilde{\mathcal{L}}_N^{\left(0\right)}\xi_t\|
\leq C_V N^{-1}m^2\|\xi_t\|\;.
\end{align}
Concerning $\mathbb{Q}^{\left(1\right)}_2$,  we start writing
\begin{align*}
\mathbb{Q}^{\left(1\right)}_2\xi_t
&=\frac{1}{Z_m}\frac{1}{N}\sum_{p\in\Lambda^*_+}p^2 a^*_pa_p(\mathcal{N}_+-1) b^*\left(\varphi_{1,t}\right)\ldots b^*\left(\varphi_{m,t}\right)\Omega\;.
\end{align*}
Since $\supp \varphi_j\subset B(0,\sqrt{\sigma})$, we can restrict the sum over $p$ on the same ball. Therefore, we have
\begin{equation}\label{eq:er2}
\begin{split}
\|\mathbb{Q}^{\left(1\right)}_2\xi_t\|
&\leq\frac{\sigma}{Z_m}\frac{1}{N}\|\cN_+(\mathcal{N}_++1) b^*\left(\varphi_{1,t}\right)\ldots b^*\left(\varphi_{m,t}\right)\Omega\|\\
&\leq \frac{C_\sigma}{Z_m N} \|(\mathcal{N}_++1)^2 b^*\left(\varphi_{1,t}\right)\ldots b^*\left(\varphi_{m,t}\right)\Omega\| \leq C_\sigma N^{-1} m^2 \|\xi_t\|\;.
\end{split}
\end{equation}
To bound $\mathbb{Q}_2^{\left(2\right)}$, we use that $V\in L^1(\mathbb{R}^3)$, so that, for any $\zeta\in\mathscr{F}^{\leq N}_+$,
\begin{equation}\label{eq:er3}
\begin{split}
\|\mathbb{Q}^{\left(2\right)}_2\zeta\|
&=\frac{1}{N}\Big\|\sum_{p\in\Lambda^*_+}\widehat{V}(p) a^*_pa_p \zeta\Big\|\leq\frac{\lVert V\rVert_{L^1}}{N}\|\cN_+ \zeta\|\;,
\end{split}
\end{equation}
which means that
\begin{equation}\label{eq:er33}
\begin{split}
\|\mathbb{Q}^{\left(2\right)}_2\xi_t\|\leq \frac{C_V m}{N}\|\xi_t\|\;.
\end{split}
\end{equation}
We now consider $\mathbb{Q}_2^{\left(3\right)}$.
Recalling the definition of $P_L$ in~\eqref{eq:PL}, we have that $P^c_L\backslash \{0\}=\setst{p\in\Lambda^*_+}{|p|> N^\alpha}$. This means that the terms proportional to $b^*_pb_p$ and $b_pb_{-p}$ vanish when acting on $\xi_t$, since $\supp \varphi_j\subset B(0,\sqrt{\sigma})$. For the contribution proportional to $b^*_pb^*_{-p}$ we use that $\widehat{V}\left(p\right)$ decays at infinity (at least as $1/|p|^{3+\varepsilon}$), so that
\begin{align}\label{eq:er4}
\|\mathbb{Q}_2^{\left(3\right)}\xi_t\|& \leq\frac{1}{2}    \sum_{p\in P_L^c\backslash \{0\}}\lvert\widehat{V}(p)\rvert\|b^*_pb^*_{-p}\xi_t\|\leq\frac{C_V}{N^{\varepsilon\alpha}}    \|(\cN_++1)\xi_t\|\leq \frac{C_V m}{\varepsilon N^{\varepsilon\alpha}}\|\xi_t\|\;.
\end{align}

We now consider $\mathcal{L}_N^{\left(3\right)}$ as defined in \eqref{eq:LNterms}. We use again that $\supp \varphi_j\subset B(0,\sqrt{\sigma})$ to restrict the sum over $q$ to the set $ B(0,2\sqrt{\sigma})$. We obtain
\begin{equation}\label{eq:er5}
\begin{split}
 \|\mathcal{L}_N^{\left(3\right)}\xi_t\|&\leq\frac{1}{\sqrt{N}}\sum_{\substack{p,q\in\Lambda^*_+\\p+q\neq 0}}\lvert \widehat{V}(p) \rvert\|\left(b^*_{p+q}a^*_{-p}a_q+a^*_qa_{-p}b_{p+q}\right)\xi_t\| \\
 &\leq \frac{C\sigma^{3/2}}{\sqrt{N}}\sum_{\substack{p\in\Lambda^*_+}}\lvert \widehat{V}(p) \rvert\|(\cN_++1)^{3/2}\xi_t\| \leq \frac{C_{V,\sigma} m^{3/2}}{N^{1/2}}\|\xi_t\|\;,
\end{split}
\end{equation}
where we used that $\sum_{\substack{p\in\Lambda^*_+}}\lvert \widehat{V}(p) \rvert\leq C_V$. Similarly, for $\mathcal{L}_N^{\left(4\right)}$ as defined in \eqref{eq:LNterms} we estimate
\begin{equation}\label{eq:er6}
\begin{split}
\|\mathcal{L}_N^{\left(4\right)}\xi_t\|& \leq \frac{1}{2N}\sum_{\substack{p,q\in\Lambda^*_+\\r\in\Lambda^*\\r\neq-p,-q}}\lvert \widehat{V}(r) \rvert\|a^*_{p+r}a^*_qa_pa_{q+r}\xi_t\|\\& \leq \frac{C\sigma^3}{2N}\sum_{\substack{r\in\Lambda^*}}\lvert \widehat{V}(r) \rvert\|(\cN_++1)^{2}\xi_t\|  \leq \frac{C_{V,\sigma} m^2}{N}\|\xi_t\|\;.
\end{split}
\end{equation}
The estimates \eqref{eq:er11}, \eqref{eq:er2}, \eqref{eq:er33}, \eqref{eq:er4}, \eqref{eq:er5} and \eqref{eq:er6}, combined with Lemma~\ref{stat:xit-normalized}, imply the bound in \eqref{eq:estELN} for $\norm{\mathcal{E}_{\mathcal{L}_N}\xi_t}$.

We now prove the bound for $\norm{\mathcal{E}_{\mathcal{L}_N}T\xi_t}$. By Lemma \ref{lm:Ngrow}, the estimates \eqref{eq:er1} and \eqref{eq:er3} show that
\begin{equation}\label{eq:er11Ter33T}
\|\widetilde{\mathcal{L}}_N^{\left(0\right)}T\xi_t\|
\leq C_V N^{-1}m^2\|\xi_t\|\quad\text{and}\quad
\|\mathbb{Q}^{\left(2\right)}_2T\xi_t\|\leq C_{V}N^{-1} m\|\xi_t\| \;.
\end{equation}
Since the sum in $\mathbb{Q}_2^{\left(3\right)}$ is restricted to momenta in $P_L^c\backslash \{0\}$, the argument leading to \eqref{eq:er4} applies to $T\xi_t$ and leads, by Lemma \ref{lm:Ngrow}, to
\begin{align}\label{eq:er4T}
\|\mathbb{Q}_2^{\left(3\right)}T\xi_t\|& \leq\frac{C_Vm}{\varepsilon N^{\varepsilon\alpha}}\|\xi_t\|\;.
\end{align}
Therefore, we only need to bound $\mathbb{Q}^{\left(1\right)}_2$, $\mathcal{L}_N^{\left(3\right)}$, and $\mathcal{L}_N^{\left(4\right)}$ on $T\xi_t$. We define $\mathbb{\widetilde Q}=(\mathcal{K}+1)(\cN_++1)$, where $\mathcal{K}$ denotes the kinetic energy operator defined by $\mathcal{K}=\sum_{q\in\Lambda^*_+}q^2a^*_qa_q$.  For $s\in[0,1]$, we define $T_s=\exp{\left(s B\right)}$, with $B$ the antisymmetric operator defined in \eqref{eq:bogolyubov-kernel}. We get
\begin{equation}\label{eq:commutator-argument}
    \frac{\diff}{\diff s}\norm{\mathbb{\widetilde Q}T_s \zeta}^2
    =\scalprod{T_s \zeta}{\left[\mathbb{\widetilde Q}^2,B\right]T_s \zeta}
    =2\mathrm{Re}\scalprod{T_s \zeta}{\mathbb{\widetilde Q}\left[\mathbb{\widetilde Q},B\right]T_s \zeta} \;,
\end{equation}
for any $\zeta\in\mathscr{F}^{\leq N}_+$.
We have
\begin{align*}
        \left[\mathbb{\widetilde Q},B\right] & =\sum_{p\in\Lambda^*_+}\Big[p^2\eta_p b^*_pb^*_{-p}\left(\mathcal{N}_++1\right)+\eta_p(\mathcal{K}+1) b^*_p b^*_{-p}\Big]+\hc \;.
\end{align*}
We then have
\begin{equation}\label{eq:qKt}
 \begin{split}
\frac{\diff}{\diff s}\norm{\mathbb{\widetilde Q}T_s \zeta}^2  =2\sum_{p\in\Lambda^*_+}&\eta_p\,\mathrm{Re}\,\langle\mathbb{\widetilde Q}T_s \zeta, \Big(p^2b^*_pb^*_{-p}\left(\mathcal{N}_++3\right)+p^2b_pb_{-p}\left(\mathcal{N}_+-1\right)\\
& \quad + (b^*_pb^*_{-p}+b_pb_{-p})(\mathcal{K}+1)\Big)T_s \zeta\rangle
= \mathrm{Re}\sum_{k=1}^4\mathrm{I}_k \;,
 \end{split}
\end{equation}
We estimate $\text{I}_1$ as
\begin{align*}
\abs{\mathrm{I}_1}&\leq2\sum_{p\in\Lambda^*_+}p^2|\eta_p|\big|\langle\mathbb{\widetilde Q}T_s\zeta, b^*_pb^*_{-p}\left(\mathcal{N}_++3\right)T_s \zeta\rangle\big|\\
&\leq C_V\sum_{p\in\Lambda^*_+}|\widehat{V}(p)|\|\mathbb{\widetilde Q}T_s \zeta\|\| \left(\mathcal{N}_++1\right)^2T_s \zeta\|\leq C_V\|\mathbb{\widetilde Q}T_s \zeta\|^2 \;,
\end{align*}
where in the last step we used that $\mathcal{N}_+\leq \mathcal{K}$.
A similar estimate holds also for $\mathrm{I}_2$. For  $\text{I}_3$, using \eqref{eq:etaL1}, we have
\begin{align*}
\abs{\mathrm{I}_3}&\leq2\sum_{p\in\Lambda^*_+}|\eta_p|\big|\langle (\cN_++1)^{-1}b_pb_{-p}\mathbb{\widetilde Q}T_s \zeta,(\cN_++1)(\mathcal{K}+1)T_s\zeta\rangle\big|\\
&\leq C\|\eta\|_{L^2}\|\mathbb{\widetilde Q}T_s \zeta\|\| (\cN_++1)(\mathcal{K}+1)T_s \zeta\|  \leq C_V\|\mathbb{\widetilde Q}T_s \zeta\|^2 \;.
\end{align*}
A similar argument holds for $\mathrm{I}_4$. Finally one gets
\begin{equation*}
        \frac{\diff}{\diff s}\norm{\mathbb{\widetilde Q} T_s \zeta}\leq C_V \norm{\mathbb{\widetilde Q}T_s \zeta}\;,
\end{equation*}
which implies, by Grönwall's lemma and setting $\zeta=\xi_t$, that
\begin{equation}\label{eq:Q21}
\|\mathbb{Q}^{\left(1\right)}_2 T_s \xi_t\|\leq \frac{1}{N}\|\mathbb{\widetilde Q} T_s \xi_t\|\leq \frac{C_V }{N} \|\mathbb{\widetilde Q}\xi_t\|\leq \frac{C_{V,\sigma} m^2}{N}\|\xi_t\| \;.
\end{equation}

Focusing on $\mathcal{L}_N^{\left(3\right)}$, using the same argument as in \eqref{eq:commutator-argument}, we get
\begin{equation*}
    \frac{\diff}{\diff s}\norm{\mathcal{L}_N^{\left(3\right)}T_s \zeta}^2
    =2\mathrm{Re}\scalprod{T_s \zeta}{\mathcal{L}_N^{\left(3\right)}\left[\mathcal{L}_N^{\left(3\right)},B\right]T_s \zeta} \;,
\end{equation*}
for any $\zeta\in\mathscr{F}^{\leq N}_+$. We have
\begin{align*}
    \left[\mathcal{L}_N^{(3)},B\right]&=\sum_{r\in\Lambda^*_+}\eta_r\Big(\frac{1}{\sqrt{N}}\sum_{\substack{p\in\Lambda^*_+\\p+r\neq 0}}\widehat{V}\left(p\right)b^*_{p+r}b^*_{-p}b^*_{-r}+\frac{1}{\sqrt{N}}\sum_{\substack{p\in\Lambda^*_+\\p+r\neq 0}}\widehat{V}\left(r\right)b^*_rb^*_pb_{p+r}\\
    &\hspace{2cm}-\frac{1}{N^{3/2}}\sum_{\substack{p,q\in\Lambda^*_+\\p+q\neq 0}}\widehat{V}\left(p\right)a^*_qa_{-p}b^*_ra^*_{-r}a_{p+q}\\
    &\hspace{2cm}+\frac{1}{\sqrt{N}}\sum_{\substack{p\in\Lambda^*_+\\p-r\neq 0}}\widehat{V}\left(r-p\right)a^*_pa_{p-r}\big(1-\mathcal{N}_+/N\big)b^*_{-r}\Big)+\text{h.c.}\;.
    \end{align*}
We write therefore
\begin{align*}
    \frac{\diff}{\diff s}\norm{\mathcal{L}_N^{\left(3\right)}T_s \zeta}^2
    = & \sum_{k=1}^4 \left(\mathrm{K}_k+\widetilde{\mathrm{K}}_k\right)\;,
\end{align*}
where the $\widetilde{\mathrm{K}}_k$-terms stem from the Hermitian conjugates in the commutator. We estimate
\begin{align*}
    \abs{\mathrm{K}_1}&\leq \frac{2}{\sqrt{N}}\sum_{\substack{p,r\in\Lambda^*_+\\p+r\neq 0}} \lvert \widehat{V}(p) \rvert|\eta_r|\|\mathcal{L}_N^{(3)}T_s\zeta \| \|b^*_{p+r} b^*_{-p}b^*_{-r}T_s\zeta\|\leq \frac{C_V}{\sqrt{N}}\|\mathcal{L}_N^{(3)}T_s\zeta \| \|\left(\mathcal{N}_++1\right)^{3/2}\zeta\|\;,
\end{align*}
where we used \eqref{eq:etaL1}, the hypothesis on $V$, and  Lemma~\ref{lm:Ngrow}. An analogous estimate holds also for $\widetilde{\mathrm{K}}_1$. Similarly, for  $\text{K}_2$, we have
\begin{align*}
    \abs{\mathrm{K}_2}&\leq \frac{2}{\sqrt{N}}\sum_{\substack{p,r\in\Lambda^*_+\\p+r\neq 0}} \lvert \widehat{V}(r) \rvert|\eta_r|\|(\cN_++1)^{-1}b_{r} b_{p}\mathcal{L}_N^{(3)}T_s\zeta \| \|(\cN_++1)b_{p+r}T_s\zeta\|\\&\leq \frac{C_V}{\sqrt{N}}\|\mathcal{L}_N^{(3)}T_s\zeta \| \|\left(\mathcal{N}_++1\right)^{3/2}\zeta\|\;,
\end{align*}
where we used \eqref{eq:bogolyubov-kernel-bound}, the hypothesis on $V$, and  Lemma~\ref{lm:Ngrow}. A similar estimate holds for $\widetilde{\mathrm{K}}_2$. We proceed to estimate $\text{K}_3$ as
\begin{align*}
    \abs{\mathrm{K}_3}&\leq \frac{2}{N^{3/2}}\sum_{\substack{p,q,r\in\Lambda^*_+\\p+q\neq 0}}|\widehat{V}(p)||\eta_r||\big\langle \left(\mathcal{N}_++1\right)^{-1/2}a_q \mathcal{L}_N^{(3)}T_s\zeta,\left(\mathcal{N}_++1\right)^{1/2}a_{-p}b^*_r a^*_{-r}a_{p+q}T_s\zeta \big\rangle|\\
    &\leq \frac{2}{N^{3/2}}\sum_{p,r\in\Lambda^*_+}|\widehat{V}(p)||\eta_r| \|\mathcal{L}_N^{(3)}T_s\zeta\| \|\left(\mathcal{N}_++1\right)^{5/2}T_s\zeta\|\\
    &\leq \frac{C_V}{N^{3/2}}\|\mathcal{L}_N^{(3)}T_s\zeta\| \|\left(\mathcal{N}_++1\right)^{5/2} \zeta\|\;,
\end{align*}
where we used again \eqref{eq:etaL1}, the hypothesis on $V$ and  Lemma~\ref{lm:Ngrow}. An analogous estimate holds also for $\widetilde{\mathrm{K}}_3$. At last, $\mathrm{K}_4$ remains to be studied. We estimate it as follows:
\begin{align}\label{eq:K4-first}
    |\mathrm{K}_4|&\leq \frac{2}{\sqrt{N}}\sum_{\substack{p,r\in \Lambda^*_+\\p-r\neq 0}} |\widehat{V}(r-p)||\eta_r||\big\langle\mathcal{L}_N^{(3)}T_s\zeta,a^*_pa_{p-r}b^*_{-r}T_s\zeta\big\rangle|\\
    &\label{eq:K4-second}\quad+\frac{2}{N^{3/2}}\sum_{\substack{p,r\in \Lambda^*_+\\p-r\neq 0}} |\widehat{V}(r-p)||\eta_r||\big\langle\mathcal{L}_N^{(3)}T_s\zeta,a^*_pa_{p-r}b^*_{-r}(\mathcal{N}_++1) T_s\zeta\big\rangle|\;.
\end{align}
The first term on the right-hand side can be estimated as
\begin{align*}
    \eqref{eq:K4-first}&\leq \frac{2}{\sqrt{N}}\sum_{p,r\in \Lambda^*_+} |\widehat{V}(r-p)||\eta_r|\|\mathcal{L}_N^{(3)}T_s\zeta\|\|\left(\mathcal{N}_++1\right)^{3/2}T_s\zeta\|\\
    &\leq \frac{C_V}{\sqrt{N}}\|\mathcal{L}_N^{(3)}T_s\zeta\|\|\left(\mathcal{N}_++1\right)^{3/2}\zeta\|\;,
\end{align*}
where we used Young's convolution inequality and Lemma~\ref{lm:Ngrow}. Similarly, one can bound \eqref{eq:K4-second} as well as $\widetilde{\mathrm{K}}_4$.

Finally, we have
\begin{equation*}
        \abs{\frac{\diff}{\diff s}\norm{\mathcal{L}_N^{\left(3\right)} T_s \zeta}^2}\leq \frac{C_V}{\sqrt{N}} \norm{\mathcal{L}_N^{\left(3\right)}T_s \zeta}\left(\|\left(\mathcal{N}_++1\right)^{3/2}\zeta\|+\frac{1}{N}\|\left(\mathcal{N}_++1\right)^{5/2}\zeta\|\right) \;.
\end{equation*}
Integrating over $s\in\left[0,1\right]$, we get
\begin{equation*}
|\|\mathcal{L}_N^{\left(3\right)} T_s \zeta\|-\|\mathcal{L}_N^{\left(3\right)} \zeta\||\leq \frac{C_V}{\sqrt{N}} \left(\|\left(\mathcal{N}_++1\right)^{3/2}\zeta\|+\frac{1}{N}\|\left(\mathcal{N}_++1\right)^{5/2}\zeta\|\right) \;.
\end{equation*}
Taking $\zeta=\xi_t$ and using \eqref{eq:er5} we obtain the desired inequality:
\begin{equation}\label{eq:L3T}
\|\mathcal{L}_N^{\left(3\right)} T_s \xi_t\|\leq\frac{C_{V,\sigma} m^{3/2}}{\sqrt{N}}\|\xi_t\| \;.
\end{equation}

Focusing on $\mathcal{L}_N^{\left(4\right)}$, using the same argument as in \eqref{eq:commutator-argument}, we get
\begin{equation*}
    \frac{\diff}{\diff s}\norm{\mathcal{L}_N^{\left(4\right)}T_s \zeta}^2
    =2\mathrm{Re}\scalprod{T_s \zeta}{\mathcal{L}_N^{\left(4\right)}\left[\mathcal{L}_N^{\left(4\right)},B\right]T_s \zeta} \;,
\end{equation*}
for any $\zeta\in\mathscr{F}^{\leq N}_+$. We have
\begin{equation*}
\begin{split}
      \left[\mathcal{L}_N^{\left(4\right)},B\right] & =\frac{1}{2N}\sum_{\substack{p,q\in\Lambda^*_+,  \,r\in\Lambda^*\\r\neq-p,-q}}\widehat{V}\left(r\right)\eta_{q+r}b^*_{-q-r}b^*_{p+r}a^*_qa_p \\
    & \quad +\frac{1}{2N}\sum_{\substack{p,q\in\Lambda^*_+,\,r\in\Lambda^*\\r\neq-p,-q}}\widehat{V}\left(r\right)\eta_p b^*_{-p}b^*_{p+r}a^*_qa_{q+r}\\
    & \quad +\frac{1}{2N}\sum_{\substack{q\in\Lambda^*_+,\, r\in\Lambda^*\\r\neq-q}}\widehat{V}\left(r\right)\eta_{q+r}b^*_qb^*_{-q}+\text{h.c.}\;,
\end{split}
\end{equation*}
which implies
\begin{align*}
    \frac{\diff}{\diff s}\norm{\mathcal{L}_N^{\left(4\right)}T_s \zeta}^2
    & =\frac{1}{N}\mathrm{Re}\big\langle\mathcal{L}_N^{\left(4\right)}T_s\zeta,\Big(\sum_{\substack{p,q\in\Lambda^*_+,\,r\in\Lambda^*\\r\neq-p,-q}}\!\!\!\!\!\!\widehat{V}\left(r\right)\eta_{q+r}b^*_{-q-r}b^*_{p+r}a^*_qa_p\\
    & \quad +\sum_{\substack{p,q\in\Lambda^*_+,\,r\in\Lambda^*\\r\neq-p,-q}}\!\!\!\!\!\!\widehat{V}\left(r\right)\eta_p b^*_{-p}b^*_{p+r}a^*_qa_{q+r}+\sum_{\substack{q\in\Lambda^*_+,\,r\in\Lambda^*\\r\neq-q}}\!\!\!\!\!\!\widehat{V}\left(r\right)\eta_{q+r}b^*_qb^*_{-q}+\text{h.c.}\Big)T_s\zeta\big\rangle\\
    &= \sum_{k=1}^3\left(\mathrm{L}_k+ \widetilde{\mathrm{L}}_k\right)\;,
\end{align*}
where the $\widetilde{\mathrm{L}}_k$-terms stem from the Hermitian conjugates in the commutator. We estimate $\text{L}_1$ as
\begin{align*}
    \abs{\mathrm{L}_1}&\leq \frac{1}{N}\sum_{\substack{p,q\in\Lambda^*_+\\r\in\Lambda^*\\r\neq-p,-q}}\lvert \widehat{V}(r) \rvert|\eta_{q+r}||\big\langle \left(\mathcal{N}_++1\right)^{-1}b_{-q-r} b_{p+r}\mathcal{L}_N^{\left(4\right)}T_s\zeta,\left(\mathcal{N}_++1\right) a^*_qa_p T_s\zeta\big\rangle|\\
    &\leq \frac{1}{N}\sum_{\substack{q\in\Lambda^*_+\\r\in\Lambda^*\\r\neq-q}}\lvert \widehat{V}(r) \rvert|\eta_{q+r}|\|\mathcal{L}_N^{\left(4\right)}T_s\zeta\|\|\left(\mathcal{N}_++1\right)^2 T_s\zeta\|
    \leq \frac{C_V}{N}\|\mathcal{L}_N^{\left(4\right)}T_s\zeta\|\|\left(\mathcal{N}_++1\right)^2 \zeta\|\;,
\end{align*}
where in the last step we used Lemma~\ref{lm:Ngrow}. A similar estimate holds also for $\widetilde{\mathrm{L}}_1$. For  $\text{L}_2$, we have
\begin{align*}
    \abs{\mathrm{L}_2}&\leq \frac{1}{N}\sum_{\substack{p,q\in\Lambda^*_+\\r\in\Lambda^*\\r\neq-p,-q}}\lvert \widehat{V}(r) \rvert|\eta_{p}|\big|\big\langle \left(\mathcal{N}_++1\right)^{-3/2}a_qb_{p+r} b_{-p}\mathcal{L}_N^{\left(4\right)}T_s\zeta,\left(\mathcal{N}_++1\right)^{3/2} a_{q+r} T_s\zeta\big\rangle\big|\\
    &\leq \frac{1}{N}\sum_{\substack{p\in\Lambda^*_+\\r\in\Lambda^*}}\lvert \widehat{V}(r) \rvert|\eta_{p}|\|\mathcal{L}_N^{\left(4\right)}T_s\zeta\|\|\left(\mathcal{N}_++1\right)^2 T_s\zeta\|  \leq \frac{C_V}{N}\|\mathcal{L}_N^{\left(4\right)}T_s\zeta\|\|\left(\mathcal{N}_++1\right)^2 \zeta\|\;,
\end{align*}
where in the last step we used again Lemma~\ref{lm:Ngrow}. A similar argument holds for $\widetilde{\mathrm{L}}_2$. For  $\text{L}_3$, we have
\begin{align*}
    \abs{\mathrm{L}_3}&\leq \frac{1}{N}\sum_{\substack{q\in\Lambda^*_+\\r\in\Lambda^*\\r\neq -q}}|\widehat{V}(r)||\eta_{q+r}|\big|\big\langle\mathcal{L}_N^{\left(4\right)}T_s\zeta,b^*_qb^*_{-q}T_s\zeta\big\rangle\big|\\
    &\leq \frac{C_V}{N}\|\mathcal{L}_N^{\left(4\right)}T_s\zeta\|\|\left(\mathcal{N}_++1\right) T_s\zeta\|  \leq \frac{C_V}{N}\|\mathcal{L}_N^{\left(4\right)}T_s\zeta\|\|\left(\mathcal{N}_++1\right) \zeta\|\;,
\end{align*}
where in the last step we used again Lemma~\ref{lm:Ngrow}, and a similar estimate holds for $\widetilde{\mathrm{L}}_3$.

Finally, we have
\begin{equation*}
        \abs{\frac{\diff}{\diff s}\norm{\mathcal{L}_N^{\left(4\right)} T_s \zeta}^2}\leq \frac{C_V}{N} \|\mathcal{L}_N^{\left(4\right)}T_s \zeta\|\left(\|\left(\mathcal{N}_++1\right) \zeta\|+\|\left(\mathcal{N}_++1\right)^2\zeta\|\right) \;.
\end{equation*}
Integrating over $s\in\left[0,1\right]$, we get
\begin{equation*}
|\|\mathcal{L}_N^{\left(4\right)} T_s \zeta\|-\|\mathcal{L}_N^{\left(4\right)}\zeta\||\leq \frac{C_V}{N}\left(\|\left(\mathcal{N}_++1\right) \zeta\|+\|\left(\mathcal{N}_++1\right)^2 \zeta\|\right)\;,
\end{equation*}
which means
\begin{equation}\label{eq:L4T}
\|\mathcal{L}_N^{\left(4\right)} T_s \xi_t\|\leq\frac{C_{V,\sigma} m^{2}}{N}\|\xi_t\| \;.
\end{equation}
From \eqref{eq:er11Ter33T}, \eqref{eq:er4T}, \eqref{eq:Q21}, \eqref{eq:L3T}, \eqref{eq:L4T} and Lemma~\ref{stat:xit-normalized} the estimate for $\norm{\mathcal{E}_{\mathcal{L}_N}T\xi_t}$ in \eqref{eq:estELN} follows.
\end{proof}

\section{Proof of Proposition \ref{prop:errorMN}}\label{sec:proof-of-prop}

\begin{proof}[Proof of Proposition \ref{prop:errorMN}]
Recall the definition in \eqref{eq:MN}. By Proposition~\ref{prop:error1}, we have that
\begin{align}\label{eq:LM}
\mathcal{M}_N&=T^*\mathcal{L}_N T=\frac{N-1}{2}\widehat{V}(0)+T^*\mathcal{Q}_NT+T^*\mathcal{E}_{\cL_N}T\;,
\end{align}
with  $\mathcal{Q}_N$ defined in \eqref{eq:QPL} and $\mathcal{E}_{\cL_N}$ satisfying \eqref{eq:estELN}.
We now consider $T^*\mathcal{Q}_NT$ in \eqref{eq:LM}. With \eqref{eq:etaMF} and \eqref{eq:genbT}, we obtain (see \cite[Section 5]{BocBreCenSch-2019-AM})
\begin{equation}\label{eq:bdiag}
    T^*\mathcal{Q}_NT=-\frac{1}{2}\sum_{p\in P_L}\left(p^2+\widehat{V}\left(p\right)-\sqrt{p^4+2\widehat{V}\left(p\right)p^2}\right)+\sum_{p\in P_L}\sqrt{p^4+2\widehat V(p)p^2}b^*_pb_p+\delta_N\;,
\end{equation}
with
\begin{equation}
\begin{split}
    \delta_N & =\mathbb{E}_1+\mathbb{E}_2 +\mathbb{E}_3+\mathbb{E}_4+\mathbb{E}_5\;,
\end{split}
\end{equation}
where
\begin{equation}\label{eq:errorDiag}
\begin{split}
\mathbb{E}_1& =\sum_{p\in P_L}(p^2+\widehat V(p))D^*_pT^*b_pT\;,\\
\mathbb{E}_2& =\sum_{p\in P_L}(p^2+\widehat V(p))[\gamma_pb^*_p+\sigma_pb_{-p}]D_p\;,\\
\mathbb{E}_3& =\frac{1}{2}\sum_{p\in P_L}\widehat V(p)[\gamma_pb^*_p+\sigma_pb_{-p}]D^*_{-p}+\hc\;,\\
\mathbb{E}_4& =\frac{1}{2}\sum_{p\in P_L}\widehat V(p)D^*_{p}T^*b^*_{-p}T+\hc\;,\\
\mathbb{E}_5&=\frac{1}{2N}\sum_{p\in P_L}\left(p^2+\widehat{V}(p)-\sqrt{p^4+2\widehat{V}(p)p^2}\right)\left(\mathcal{N}_++a^*_pa_p\right)\;,
\end{split}
\end{equation}
and we denoted $\gamma_p=\cosh(\eta_p)$ and $\sigma_p=\sinh(\eta_p)$.
We now estimate $\delta_N$. Writing $T_s =e^{sB}$, we have
\begin{equation}\nonumber
\begin{split}
T^*b_pT=b_p + \int_0^1 \diff s\,T_s^*[b_p,B]T_s\;.
\end{split}
\end{equation}
Computing the commutator, we obtain
\begin{equation}\label{eq:oneIter}
\begin{split}
T^*b_pT=b_p+\eta_p\int_0^1 \diff s\, T_s^*\Big(1-\frac{\cN_+}{N}\Big)b^*_{-p}T_s -\frac{1}{N}\int_0^1 \diff s\, T_s^*\Big(\sum_{q\in\Lambda^*_+}\eta_qb^*_qa^*_{-q}a_p\Big)T_s\;.
\end{split}
\end{equation}
Iterating these calculations (see \cite[Lemma 5.3]{BocBreCenSch-2019-AM}) leads to \eqref{eq:genbT}
with $D_p$ given by
\begin{equation}\label{eq:Dp-def}
\begin{split}
D_{p} = \; & \sum_{n \geq 0}  \int_0^1 d s_1 \ldots \int_0^{s_{2n}} d s_{2n+1}\\
& \hskip 1.5cm \times e^{-s_{2n+1}B(\eta)} \Big(- \eta_{p}^{2n+1} \, \frac{\cN_+}{N} b^*_{-p} - \frac 1 N \,\eta_{p}^{2n} \sum_{q \in \Lambda^*_+} b^*_q a^*_{-q} a_p \eta_{q} \Big) e^{s_{2n+1}B(\eta)} \\
& + \; \sum_{n \geq 1}  \int_0^1 d s_1 \ldots \int_0^{s_{2n-1}} d s_{2n}\\
& \hskip 1.5cm \times e^{-s_{2n}B(\eta)} \Big(- \eta_{p}^{2n} b_p\, \frac{\cN_+}{N}  - \frac 1 N \,\eta_{p}^{2n-1} \sum_{q \in \Lambda^*_+} a^*_{-p} a_{-q} b_q \eta_{q} \Big) e^{s_{2n}B(\eta)}\;,
\end{split}
\end{equation}
and satisfying, for any $\zeta\in\cF$,
\begin{equation}\label{eq:esTD}
\begin{split}
\| (\cN_++1)^{n/2} D_{p} \zeta \| \leq  \frac{C_V}{N} |\eta_{p}|  \|(\cN_++1)^{(n+3)/2} \zeta \| + \frac{C_V}{N} \int_0^1ds\; \| a_p (\cN_+ +1)^{(n+2)/2} T_s \zeta \|
\end{split}
\end{equation}
and
\begin{equation}\label{eq:esTDstar}
\begin{split}
\| (\cN_++1)^{n/2} D^*_{p} \zeta \| \leq  \frac{C_V}{N}  \|(\cN_++1)^{(n+3)/2} \zeta \|\;,
\end{split}
\end{equation}
which follow from \eqref{eq:etaL1} and Lemma~\ref{lm:Ngrow} (note that the constants in \eqref{eq:esTD} and \eqref{eq:esTDstar} do not depend on the choice of $P_L$).
This means that, for $\xi_t$ as in \eqref{eq:evolved-initial-state},
\begin{equation*}
\begin{split}
\|\mathbb{E}_1\xi_t\|& \leq\sum_{p\in P_L}(p^2+|\widehat V(p)|)\|D^*_pT^*b_pT\xi_t\|\leq \frac{C_V}{N}\sum_{p\in P_L}(p^2+|\widehat V(p)|)\|(\cN_++1)^{3/2}T^*b_pT\xi_t\|\;.
\end{split}
\end{equation*}
By means of~\eqref{eq:oneIter}, \eqref{eq:etaL1}, the unitarity of $T_s$ and Lemma~\ref{lm:Ngrow}, we get
\begin{equation*}
\begin{split}
\|\mathbb{E}_1\xi_t\|& \leq \frac{C_V}{N}\sum_{p\in\Lambda^*_+\cap B\left(0,\sqrt{\sigma}\right)}(p^2+|\widehat V(p)|)\|(\cN_++1)^{3/2}b_p\xi_t\|\\
&\qquad+\frac{C_V}{N}\sum_{p\in P_L}|\eta_p|(p^2+|\widehat V(p)|)\|(\cN_++1)^{2}\xi_t\|\\
&\qquad+\frac{C_V}{N^2}\sum_{p\in P_L}(p^2+|\widehat V(p)|)\|(\cN_++1)^{3}\xi_t\|\;,
\end{split}
\end{equation*}
which leads to 
\begin{equation}\label{eq:erD1}
\begin{split}
\|\mathbb{E}_1\xi_t\|& \leq \frac{C_V\sigma^{5/2}}{N}\|(\cN_++1)^{2}\xi_t\|+\frac{C_V N^{5\alpha}}{N^2}\|(\cN_++1)^{3}\xi_t\|\;.
\end{split}
\end{equation}
Similarly, for $\mathbb{E}_2$, using \eqref{eq:esTD}, using \eqref{eq:etaL1} and estimating $|\sigma_p|\leq C_V|\eta_p|$  we have
\begin{equation}\label{eq:erD2}
\begin{split}
\|\mathbb{E}_2\xi_t\|& \leq\sum_{p\in P_L}(p^2+|\widehat V(p)|)\|\left( \gamma_pb^*_p+\sigma_pb_{-p}\right) D_p\xi_t\|\\& \leq C_V\sum_{p\in P_L}(p^2+|\widehat V(p)|)\|(\cN_++1)^{1/2}D_p\xi_t\| \leq \frac{C_VN^{5\alpha}}{N}\|(\cN_++1)^{2}\xi_t\|\;.
\end{split}
\end{equation}
For $\mathbb{E}_3$, using~\eqref{eq:esTD} and~\eqref{eq:esTDstar}, we see that
\begin{equation}\label{eq:erD3}
\begin{split}
\|\mathbb{E}_3\xi_t\|& \leq\sum_{p\in P_L}|\widehat V(p)|\|\left( \gamma_pb^*_p+\sigma_pb_{-p}\right) D^*_{-p}\xi_t\|+\sum_{p\in P_L}|\widehat V(p)|\|D_{-p}\left( \gamma_pb_p+\sigma_pb^*_{-p}\right) \xi_t\|\\
&\leq \frac{C_V}{N}\|(\cN_++1)^{2}\xi_t\|\;.
\end{split}
\end{equation}
Then, we obtain an estimate for $\mathbb{E}_4$, using \eqref{eq:esTDstar} and Lemma \ref{lm:Ngrow}, so that
\begin{equation}\label{eq:erD4}
\begin{split}
\|\mathbb{E}_4\xi_t\|& \leq\sum_{p\in P_L}\lvert\widehat{V}(p)\rvert\|D^*_{p}T^*b^*_{-p}T\xi_t\|+\sum_{p\in P_L}\lvert \widehat{V}(p)\rvert\|T^*b_{-p}TD_{p}\xi_t\|\\
&\leq\frac{C_V}{N}\|(\cN_++1)^{2}\xi_t\|\;.
\end{split}
\end{equation}
Finally, for $\mathbb{E}_5$, we get
\begin{align*}
    \|\mathbb{E}_5\xi_t\|&\leq \frac{1}{2N}\sum_{p\in P_L}\frac{|\widehat{V}(p)|^2}{\Big\vert p^2+\widehat{V}(p)+\sqrt{p^4+2\widehat{V}(p)p^2}\Big\vert}\|\left(\mathcal{N}_++a^*_pa_p\right)\xi_t\|\ \leq \frac{C_V}{N}\|\mathcal{N}_+\xi_t\|\;.
\end{align*}

By \eqref{eq:LM} and \eqref{eq:bdiag}, we see that \eqref{eq:cM} holds with $E_N$ given by \eqref{eq:phase}, $\mathbb{H}_{\textnormal{B}}$ given by \eqref{eq:diagonal-bogolyubov-hamiltonian} and $\mathcal{E}_N$  given by
\begin{equation}\label{eq:EN-error}
\begin{split}
    \mathcal{E}_N  & =T^*\mathcal{E}_{\mathcal{L}_N}T+\delta_N-\frac{1}{N}\sum_{p\in P_L}\sqrt{p^4+2\widehat{V}\left(p\right)p^2}a^*_p\mathcal{N}_+a_p\\
    & \quad+\frac{1}{2}\sum_{p\in P_L^c\backslash\{0\}}\left(p^2+\widehat{V}\left(p\right)-\sqrt{p^4+2p^2\widehat{V}\left(p\right)}\right)\;.
\end{split}
\end{equation}
The last contribution in \eqref{eq:EN-error} is bounded by $C_V\varepsilon^{-1}N^{-\alpha\varepsilon}$ since the summand behaves as $|\widehat{V}(p)|^2p^{-2}$. The second to last contribution can be bounded as \eqref{eq:er2}.
These considerations, together with the estimates in \eqref{eq:erD1}, \eqref{eq:erD2}, \eqref{eq:erD3}, \eqref{eq:erD4}, and with the bound in \eqref{eq:estELN}, lead to the bound in \eqref{eq:cEN}.
\end{proof}

\addtocontents{toc}{\protect\setcounter{tocdepth}{0}}

\section*{Acknowledgements} C.~B.\ and N.~B.\ would like to thank Dario Bambusi for useful discussions.
C.~B.\ and U.~M.\ acknowledge financial support from MUR–Italian Ministry of University and Research and Next Generation EU within PRIN 2022AKRC5P ``Interacting Quantum Systems: Topological Phenomena and Effective Theories''. C.~B. also acknowledges support of Grant PID2024-156184NB-I00 funded by
MICIU/AEI/10.13039/ 501100011033 and cofunded by the European Union. N.~B.\ and U.~M.\ acknowledge financial support from the European Union through the European Research Council’s Starting Grant \textsc{FermiMath}, grant agreement n.~101040991. Views and opinions expressed are those of the authors and do not necessarily reflect those of the European Union or the European Research Council Executive Agency. Neither the European Union nor the granting authority can be held responsible for them. The authors have also been partially supported by Gruppo Nazionale per la Fisica Matematica GNFM – INdAM.

\section*{Declarations}
The authors declare no conflict of interest.

ChatGPT 6 was used for proofreading and language editing. No AI tools were used in developing the mathematical arguments.

\section*{Data availability statement}
Data sharing is not applicable to this article as it has no associated data.

\addtocontents{toc}{\protect\setcounter{tocdepth}{2}}

\addtocontents{toc}{\protect\setcounter{tocdepth}{0}}

\addtocontents{toc}{\protect\setcounter{tocdepth}{2}}

%%% BIBLIOGRAPHY %%%
\bibliographystyle{alpha} %try "alpha", "unsrt", "plain" or "siam"
\small{\bibliography{bibliography}}

\end{document}